\documentclass[%
 reprint,
superscriptaddress,
 amsmath,amssymb,
 aps,
prl,
floatfix
]{revtex4-1}

\usepackage{lmodern}
\usepackage{graphicx}
\usepackage{bm}
\usepackage{hyperref}
\hypersetup{linktocpage,colorlinks,citecolor={blue},pdfdisplaydoctitle=true,pdfpagemode=UseOutlines,bookmarksnumbered=true}
\usepackage{mathrsfs,dsfont}
\usepackage{color}
\usepackage{cleveref}

\newcommand{\bra}[1]{\langle #1 |} 
\newcommand{\ket}[1]{| #1 \rangle } 
\newcommand{\upd}{\mathrm{d}}
\newcommand{\tr}{\mathrm{tr}}

\newcommand{\eg}[0]{\textit{e.g.} }
\newcommand{\xb}[0]{\mathbf{x}}
\newcommand{\yb}[0]{\mathbf{y}}
\newcommand{\zb}[0]{\mathbf{z}}

\newcommand{\kb}[0]{\mathbf{k}}

\newcommand{\e}[0]{\mathrm{e}}
\newcommand{\tms}[1]{\textcolor{black}{#1}}

\begin{document}

\title{Neutron star heating constraints on wave-function collapse models}

\author{Antoine Tilloy}
\email{antoine.tilloy@mpq.mpg.de}
\affiliation{Max-Planck-Institut f\"ur Quantenoptik, Hans-Kopfermann-Stra{\ss}e 1, 85748 Garching, Germany}

\author{Thomas M. Stace}
\affiliation{ARC Centre for Engineered Quantum Systems, School of Mathematics and Physics, The University of Queensland, Brisbane, QLD 4072, Australia}

\begin{abstract}
    Spontaneous wavefunction collapse models, like the Continuous Spontaneous Localization, are designed to 
    suppress macroscopic superpositions, while preserving microscopic quantum phenomena.   An observable consequence of collapse models is spontaneous heating of massive objects.  Here we calculate the  collapse-induced heating rate of astrophysical    objects, and the corresponding equilibrium temperature. We apply these results to neutron stars, the  densest phase of baryonic matter in the universe.  Stronger collapse model parameters imply greater heating, allowing us to derive competitive bounds on  model parameters using neutron star observational data, and to propose speculative  bounds based on the capabilities of current and future astronomical surveys.
\end{abstract}

\maketitle

 Collapse models, like the Continuous Spontaneous Localization (CSL) model \cite{pearle1989,ghirardi1990}, aim at solving the measurement problem of quantum mechanics through a  stochastic non-linear modification of the Schr\"odinger equation \cite{bassi2003,bassi2013review}. Such modifications have sometimes been conjectured to be caused by gravity, the most famous example being the Di\'osi-Penrose (DP) model \cite{diosi1987,penrose1996}. In general, collapse models posit an intrinsic (possibly gravitational) noise, which endogenously  collapses  superpositions of sufficiently macroscopic systems (in a particular basis), while preserving the predictions of quantum mechanics at small scales.  One notable consequence of these models is spontaneous heating of massive objects.   Neutron stars, which are  extremely dense, macroscopic quantum-limited objects, offer  a unique system on which to test this prediction.  Here, we estimate the equilibrium temperature of a neutron star radiating  heat generated from spontaneous collapse models. We find that neutron stars are competitive to constrain the parameter diagram of collapse models. Theoretically or observationally improving upper bounds for neutron star equilibrium temperatures could in principle eliminate historically proposed CSL parameter values.

\paragraph{Collapse models --} Continuous Markovian collapse models modify the Schr\"odinger equation with a non-linear noise term:
\begin{equation}\label{eq:sto}
    \partial_t \ket{\psi_t} = - \frac{i}{\hbar} H \ket{\psi_t} + F(\eta_t,\ket{\psi_t})
\end{equation}
where $\eta_t$ is a white noise  process and $F$ some function which is partially constrained by consistency conditions \cite{gisin1989,wiseman2001}, and is chosen to yield a spontaneous collapse in the position basis.

Although this stochastic description \eqref{eq:sto} of the state vector is required to understand why collapse models actually achieve their purpose and solve the measurement problem, their empirical content is fully contained in the master equation obeyed by $\rho_t = \mathds{E}\big[\ket{\psi_t}\bra{\psi_t}\big]$. For \emph{most} Markovian non-dissipative collapse models proposed so far \cite{bassi2013review}, it takes the form $\partial_t \rho_t = -\frac{i}{\hbar} [H,\rho_t] + \mathcal{D}[\hat{M}]\rho_t$ with
\begin{equation}\label{eq:master}
 \mathcal{D}[\hat{M}]\rho = - \!\int \! \upd\xb \upd\yb \, f(\xb-\yb)\! \left[\hat{M}_{r_c}(\xb),\left[\hat{M}_{r_c}(\yb),\rho\right]\right]
\end{equation}
where $f$ is a positive definite function and $\hat{M}_{r_c}(\xb)$ is a regularized mass density operator:
\begin{equation}\label{eq:M}
    \hat{M}_{r_c}(\xb)= g_{r_c} * \hat{M}(\xb) = g_{r_c} * m\, a^\dagger(\xb) a(\xb).
\end{equation}
In this expression, $m$ in the mass of the particle considered (we will consider neutrons), $a_k^\dagger(\xb), a_k(\xb)$ denote the usual (here fermionic) creation and annihilation operators, $g_{r_c}$ is a regulator which smooths the mass density over a length scale $r_c$ and ``$*$'' denotes the convolution product. Typically, the regulator function is taken to be Gaussian:
\begin{equation}
    g_{r_c}(\xb) = \ e^{-\xb^2/(2r_c^2)}/(\sqrt{2\pi r_c^2})^3.
\end{equation}
The regulator length scale has to be much larger than the Planck length and even the nucleon Compton wave-length, the usual choice being $r_c \simeq 10^{-7} \mathrm{m}$  \cite{ghirardi1986}.

The two most common continuous collapse models are the Continuous Spontaneous Localization (CSL) model and the Di\'osi-Penrose model (the latter having a heuristic link with gravity):
\begin{enumerate}
    \item The CSL model is obtained for:
    \begin{equation}
    f^\text{CSL}(\xb-\yb) = \frac{\gamma}{2 m_N^2} \times \delta(\xb-\yb)
\end{equation}
where $m_N$ is the mass of a nucleon and $\gamma$ is the collapse ``strength''. It is a rate $\times$ distance$^3$, the corresponding rate is $\lambda_{\text{CSL}}\equiv \gamma/(4\pi r_c^2)^{3/2} $ historically fixed at $\lambda_{\text{CSL}}\simeq 10^{-16}$s$^{-1}$ (the so called ``GRW'' value).
 \item The DP model is obtained for:
 \begin{equation}\label{eq:DPchoice}
     f^\text{DP}(\xb-\yb)=\frac{G}{4\hbar} \times \frac{1}{|\xb-\yb|}.
 \end{equation}
 Because the collapse strength is  fixed by the gravitational constant, there is one parameter less \footnote{Note that the factor $1/4$ in \cref{eq:DPchoice} has also been fixed to $1/8$ in the literature.}. A modern motivation for \cref{eq:DPchoice} is given by attempts at constructing models of fundamentally semiclassical gravity \cite{tilloy2016,tilloy2017principle}.
\end{enumerate}
We note that, at least at the master equation level, the regulator applied on the mass density operator can equivalently be applied on the kernel $f$:
   \begin{equation}\label{eq:masterbis}
 \mathcal{D}[\hat{M}]\rho = - \!\int \! \upd\xb \upd\yb \, f_{r_c}(\xb-\yb)\! \left[\hat{M}(\xb),\left[\hat{M}(\yb),\rho\right]\right],
\end{equation} 
with $f_{r_c}=g_{r_c}*f*g_{r_c}$.

We also note in passing that the two models we consider here are non-relativistic.  Efforts towards developing relativistic collapse models for quantum fields have shown that their construction is possible (albeit challenging, see \eg \cite{diosi1990relat,pearle1999,tumulka2006,bedingham2011,pearle2015,tilloy2017qft,juarez-aubry2018}). Here, we simply assume that such relativistic extensions can be constructed, and that, in the limit where relativistic effects are not dominant, their predictions would be similar to those of the non-relativistic CSL or DP models.

\paragraph{Spontaneous heating --}
The additional decoherence term \cref{eq:master} in the master equation does not commute with the kinetic part of the Hamiltonian, hence the expectation of the energy $\langle H \rangle_t \equiv \tr [H \rho_t]$ is no longer conserved. This spontaneous heating provides a natural test of collapse models \cite{pearle1994, bahrami2018, adler2018}. 

Recent proposals to test these models have \eg been built around ultra cold atoms \cite{laloe2014}, which may provide good platforms to obtain bounds on the parameters in the theory as the heating effect should be significant in relative terms. An alternative, which has been overlooked so far, is to consider instead maximally dense systems, exploiting the mass density dependence of the heating for all collapse models. In this respect, neutron stars  are  ideal candidates.

Neutron star cooling has been studied theoretically and observationally.  At early stages when $T_\text{star}\sim 10^9$ K, they cool by various baryonic emission processes, but at later stages, when $T_\text{star}\sim 10^6$ K or colder, the cooling is radiation dominated \cite{lat94,lat01,yak04}.  Thus, the equilibrium temperature is attained by the balance of the spontaneous collapse induced heating with Stefan-Boltzmann radiation, so is determined by the heat balance condition \mbox{$P_{\text{heat}}= P_{\text{rad}}$}, where
    \begin{equation}\label{eq:pcollapse}
        P_{\text{heat}}=\partial_t \langle H \rangle_t= \tr[ H\, \mathcal{D}[\hat M] \rho_t]
    \end{equation}
and 
    \begin{equation}\label{eq:stefan}
        P_{\text{rad}} = S \sigma T^4
    \end{equation}
where $S$ is the neutron star surface area and \mbox{$\sigma=5.6\cdot 10^{-8}\mathrm{W}\cdot \mathrm{m}^{-2}\cdot \mathrm{K}^{-4}$} is Stefan's constant.
It follows that at equilibrium $T_{\rm star}=\big(P_{\rm heat}/(S\sigma)\big)^{1/4}$.

For a system of $N$ fermions with non-relativistic Hamiltonian, one can show that the spontaneous collapse induced heating $P_\text{heat}$ is independent of the potential (which commutes with the mass density) and more surprisingly does not even depend on the quantum state. For the CSL model it reads:
\begin{equation}\label{eq:cslheating}
    P_\text{heat}^{\rm CSL} = \tr[ H\, \mathcal{D}[\hat M] \rho_t] = \frac{3\lambda \hbar^2}{4 r_c^2 m} N,
\end{equation}
where $N$ is the number of neutrons in the star. Similarly, for the DP model it reads:
\begin{equation}\label{eq:dpheating}
    P_\text{heat}^\text{DP} = \frac{G\hbar m}{8 \sqrt{\pi} r_c^3} N.
\end{equation}

\paragraph{The CSL model --}
We take the typical neutron star radius \mbox{$L\sim 10$ km} and mass \mbox{$M_{\rm star}\sim M_{\odot}\simeq 2.0 \cdot 10^{30}$ kg}, hence \mbox{$N = M_{\rm star}/m_{N} \simeq 10^{57}$} neutrons. For the values historically proposed for the CSL model, $\lambda=10^{-16}\textrm{s}$ and $r_c=10^{-7}\textrm{m}$, one finds $P_\text{heat}\sim 10^{14}\textrm{W}$. On the other hand the lowest observed temperature of an astronomical neutron star is $T^\mathrm{(obs)}=0.28$ MK  for the object  \texttt{PSR J 840-1419} \cite{keane2013}. This observed temperature corresponds to a radiative dissipation rate of $P^{\rm (obs)}_{\rm rad}\sim10^{26}$ W, well above the power that would be radiated by the CSL model. Hence, the neutron stars we can currently observe are not cold enough to straightforwardly falsify the CSL model.

Naturally, neutron stars are expected to cool down to much lower temperatures than the ones we currently manage to see directly \cite{yak04} and the bound from \texttt{PSR J 840-1419} is thus an excessively conservative one. \tms{We discuss this further below.}

\begin{figure}
\includegraphics[width=0.98\columnwidth]{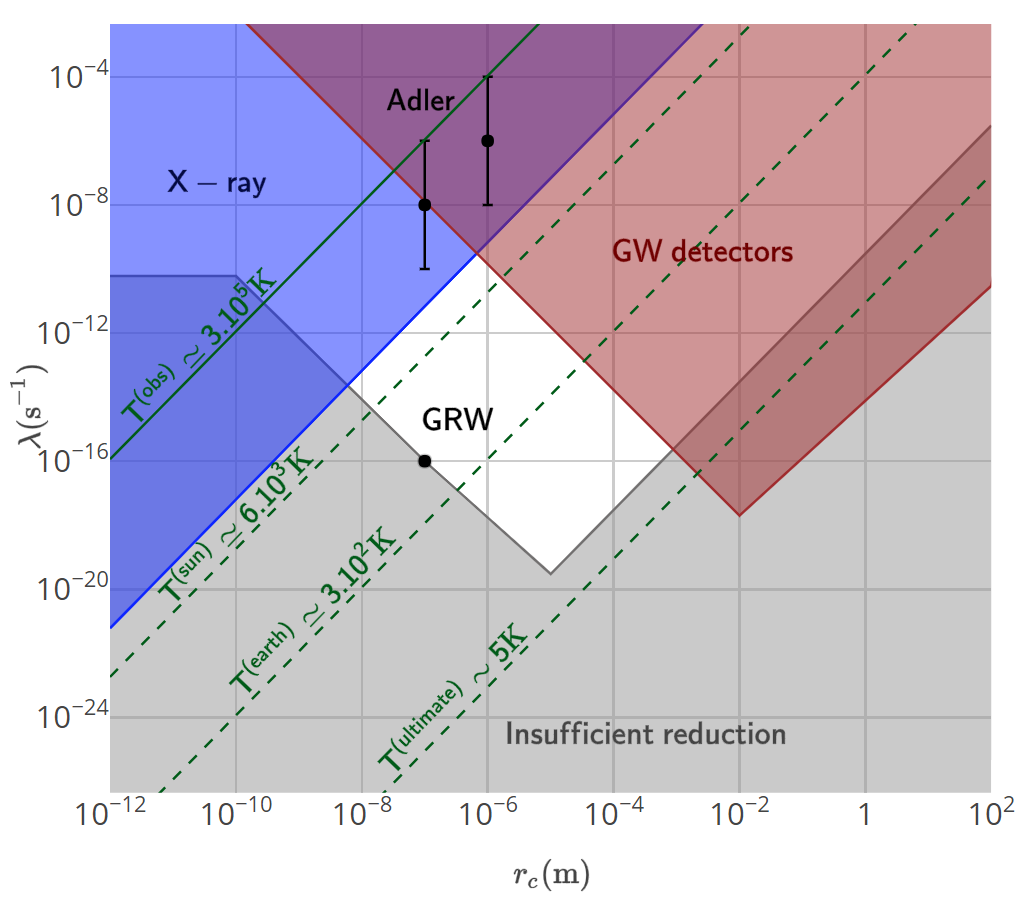}
\caption{CSL parameter diagram --
Top: Zones formerly excluded by gravitational wave detectors \cite{carlesso2016,helou2017} (red), spontaneous X-ray emission \cite{piscicchia2017} (blue), and insufficient macroscopic localization \cite{toros2017}. The value historically proposed by GRW \cite{ghirardi1986} and the range put forward by Adler \cite{adler2007} are shown with black dots. The green line delineate the upper left regions that are excluded by currently observed neutron stars (continuous line). More speculative bounds, obtained assuming various equilibrium temperatures for neutron stars, are showed in hashed green.}
\label{fig:parameters}
\end{figure}

\paragraph{The DP model --} 
Following the same reasoning as for the CSL model, we can constrain the only free parameter, the regularization length $r_c$, of the DP model using eq. \eqref{eq:dpheating}. The most conservative bound, given by \texttt{PSR J 840-1419}, yields $r_c  \gtrsim 10^{-13}$m, which excludes a regulator of the order of the neutron radius which was historically conjectured to be a possible cutoff. This lower bound is of the same order of magnitude as the current best one of $4\times10^{-14}$m yielded by constraints from gravitational wave detector data \cite{helou2017}. The bound improves  with decreasing temperatures $r_c \propto T^{-4/3}$.

\paragraph{Discussion --} 
The analysis presented in this letter makes `lumped-element' approximations that provide robust bounds on the radiated power.  For example, we have assumed that the emissivity of a neutron star is unity and that the thermal conductivity throughout the core is large enough that the star temperature is approximately uniform.  If  these assumptions are relaxed, then the core temperature may be substantially higher than the observed surface temperature.  Neutron superfluidity \cite{lat01} has been hypothesised in the core of neutron stars. This phase will have a corresponding critical temperature $T_c$, which may provide a sensitive thermometric bound on tolerable heat generation rates in the star core: superfluidity will be suppressed if the internal temperature is too high.  More generally, heat transfer models that include  realistic constitutive models for the neutron star body may thus be able to provide even more stringent bounds on  collapse model parameters than the lumped-element approximations we have adopted here.


\tms{The positive bounds established above, are based on observed temperatures of young, hot, bright neutron stars. There is a possibility for improvement in the bounds if colder neutron stars are observed, or if a large population of cold remnants can be excluded due to lack of observation, 
so we now speculate on the near-term prospects for wide survey observations.}

\tms{The Dark Energy Survey (DES) has classified a significant fraction of astronomical objects down to apparent magnitude  $m=23$ \cite{Drlica_Wagner_2018}.  The  separation of neutron stars in the vicinity of our sun is estimated to be around 10 pc \cite{NSdensity}, so the nearest neutron star is expected to be \mbox{$d\approx5$ pc} away from the Earth.  At that range, $m=23$ objects seen by DES correspond to a luminosity of \mbox{$5\times 10^{18}$ W}, and a neutron star surface temperature of \mbox{$22\times10^3$ K} (assuming a neutron star radius of 10 km). Thus, the DES should be able to see nearby, cool neutron stars.  This would put a constraint on CSL models which is roughly comparable to the constraints from spontaneous X-ray emission studies \cite{piscicchia2017}.  }

\tms{In future, the Large  Synoptic Survey Telescope (LSST)  will be able to image apparent magnitude $m=28$ objects \cite{Abell:2009aa}; at 5 pc, such objects have luminosity $5\times 10^{16}$ W, and a surface temperature of $7\times10^3 \textrm{ K}\approx T^{\rm{(sun)}}$.  Such an observation which would improve bounds on the CSL model, as shown in Fig.~\ref{fig:parameters}.}

\tms{
In the event that either DES or LSST \emph{fails} to observe such objects, it would suggest either a (surprisingly) low local density of neutron stars, or that nearby neutron stars are unobservably cold (i.e.\ $T< T^{\rm{(sun)}}$). The latter inference would further rule out  parts of the CSL parameter diagram, also shown in Fig.~\ref{fig:parameters}.}

\tms{More speculatively, we might hope to one day be able to eliminate the possibility of an equilibrium temperature like that of our own planet, $T^{(\textrm{earth})} \sim 3\times10^2 K$, which would falsify the historical GRW values by two orders of magnitude.}  

\tms{What might be the ultimate observable limit, even in principle? Neutron stars would be net thermal sources indefinitely if their minimum equilibrium temperature exceeded the cosmic microwave background (CMB) temperature.  Though this would be difficult to observe terrestrially, it does offer an intriguing limit.  Below $T^{\textrm{(ultimate)}}=5K\gtrsim T^{\textrm{(CMB)}}$, we find that the CSL parameter bounds are too low for collapse models to be effective, as shown in Fig.~\ref{fig:parameters}.}

\tms{For the DP model, upper bounds on $r_c$ can be obtained if the model is required to provide a consistent theory of fundamental semiclassical gravity \cite{tilloy2017principle}. In this context, the regulator $g_{r_c}$ affects the Newtonian potential and the $1/r^2$ law of the gravitational force breaks down for \mbox{$r\sim r_c$}. The Newtonian force is well measured for distances as short as $100\mathrm{\mu m}$ \cite{Hagedorn2015}, which provides a conservative upper bound, \mbox{$r_c\lesssim10^{-4}$ m}. Even supposing cold neutron stars of a few Kelvin, we find $r_c\gtrsim 10^{-7}$ m. Hence the range of values allowed for the DP model could not (even in principle) be closed by the temperature of neutron stars alone, and gravitational upper bounds would need to be improved in parallel. }

On the other hand, refinements and extensions of the CSL model with colored noise (cCSL) \cite{bassi2002,adler2007colored}, dissipation (dCSL) \cite{smirne2014}, or both \cite{ferialdi2012}, containing additional parameters (such as a high frequency cutoff or a temperature) are known to yield weaker heating effects. Consequently, the constraints we put forward here would be weaker for these models.

In summary, with a conservative estimate of neutron star cooling based on the currently observed coldest neutron stars, one obtains constraints on the CSL model (albeit weaker than from spontaneous X-ray emission studies) and on the DP model ($r_c\gtrsim10^{-13}$m, competitive with state of the art gravitational wave interferometer data). Improving the observational upper bound on neutron star equilibrium temperatures would yield substantial improvements. If we could measure an old, cold neutron star one could test more of the CSL parameter diagram. This  motivates a systematic survey of nearby, cold neutron stars.

\begin{acknowledgments}  We  thank Philip Pearle and Tamara Davis for helpful discussions during development of this work.  AT was supported by the Alexander von Humboldt foundation. TMS was supported by the Australian Research Council Centre of Excellence for Engineered Quantum Systems  CE170100039.
\end{acknowledgments}

\bibliographystyle{apsrev4-1}
\bibliography{main}

 \appendix
\section*{Supplementary material}
We compute the power $P_\text{heat}=\tr[ H\, \mathcal{D}[\hat M] \rho_t]$ generated by spontaneous collapse for a generic continuous Markovian non-dissipative collapse model and then evaluate the result for the CSL and DP models.

We consider a rather generic non-relativistic Hamiltonian $\hat{H}=\hat{H}_0 + \hat{V}$ for fermions with:
\begin{align}
    \hat{H}_0 &= \frac{-\hbar^2}{2m}\int \upd \xb a^\dagger(\xb) \nabla^2_\xb a(\xb)\\
    \hat{V}&= \int \upd \xb \upd \yb \, a^\dagger(\xb)a(\xb) V(\xb-\yb) a^\dagger(\yb) a(\yb),
\end{align}
where $a^\dagger(\xb)$ and $a(\xb)$ are the local anti-commuting creation / annihilation operators $\{a^\dagger(\xb),a(\yb)\}=\delta(\xb-\yb)$ and we neglect spin.
We note that $\hat{V}$ commutes with the local mass density and thus does not contribute to $P_\text{heat}$. Introducing $\hat{P}_\text{heat}$ such that $P_\text{heat} = \tr\left[\hat{P}_\text{heat}\,\rho_t\right]$ we have:
\begin{equation}
\begin{split}
    \hat{P}_\text{heat}=&\frac{\hbar^2}{2} m\int\upd \xb\upd \yb \upd \zb \,  f_{r_c}(\xb,\yb) \times\\
    &\bigg[a^\dagger(\xb) a(\xb),\Big[a^\dagger(\yb) a(\yb), a^\dagger(\zb) \nabla^2_\zb a(\zb) \Big]\bigg]
\end{split}
\end{equation}
We shall prove that:
\begin{equation}
     \hat{P}_\text{heat}= \frac{\hbar^2}{2 } m \left(-2\nabla^2_\xb f_{r_c}(\xb)|_{\xb=0} \right) \; \hat{N}
\end{equation}
where $\hat{N} = \int \upd \xb a^\dagger(\xb) a(\xb)$ is the total number of neutrons.
Let us compute the first commutator using the canonical anti-commutation relations and evaluate the $\yb$ integral:
\begin{align}
    &\int \upd \yb f_{r_c} (\xb-\yb) \Big[a^\dagger(\yb) a(\yb), a^\dagger(\zb) \nabla^2_\zb a(\zb) \Big] \nonumber \\ 
&=\int \upd \yb f_{r_c} (\xb-\yb) \big(\delta(\yb-\zb) a^\dagger(\yb)\nabla^2_\zb a(\zb) \\
    &\hskip3.4cm- \nabla^2_\zb\delta(\yb-\zb)a^\dagger(\zb) a(\yb)\big)\nonumber\\
&=a^\dagger(\zb) \left(f_{r_c}(\xb-\zb)\nabla^2_\zb a(\zb)- \nabla^2_\zb \left[f_{r_c}(\xb-\zb)a(\zb)\right]\right) \\
&=- a^\dagger(\zb)a(\zb) \nabla_\zb^2 f_{r_c}(\xb-\zb) - 2 a^\dagger (\zb) \nabla_\zb f_{r_c}(\xb-\zb) \nabla_\zb a(\zb)\label{eq:twoterms}
\end{align}
Upon insertion in the second commutator, the first term in \eqref{eq:twoterms} will vanish. Using again the canonical anti-commutation relations we get:
\begin{equation}
    \begin{split}
       [\spadesuit]\equiv& [a^\dagger(\xb) a(\xb), -2 a^\dagger(\zb) \nabla_\zb f_{r_c}(\xb-\zb) \nabla_\zb a(\zb)] \\
        =&-2 \delta(\xb-\zb) a^\dagger(\xb) \nabla_\zb f_{r_c}(\xb-\zb) \cdot \nabla_\zb a(\zb)\\
        &+2 a^\dagger(\zb) \nabla_\zb f_{r_c}(\xb-\zb) \cdot \nabla_\zb \delta(\xb-\zb) a(\xb)
    \end{split}
\end{equation}
Once integrated over $\xb$, the first term will be proportional to $\nabla_\zb f_{r_c}(\xb-\zb)|_{\xb=\zb} = 0$ by symmetry, and thus will not contribute. Using $\nabla_\zb \delta(\xb-\zb) = -\nabla_\xb \delta(\xb-\zb)$ on the second term we get:
\begin{equation}
\begin{split}
    \int \!\! \upd \xb [\spadesuit]&=-2 \int\!\! \upd \xb a^\dagger(\zb) \nabla_\zb f_{r_c}(\xb-\zb) \cdot \nabla_\xb \delta(\xb-\zb) a(\xb)\\
    &= 2 \int\!\! \upd \xb \delta(\xb-\zb) a^\dagger(\zb) \left[\nabla_\xb\cdot \nabla_\zb f_{r_c}(\xb-\zb) a(\xb)\right]  \\
    &=2 \Big[a^\dagger(\zb)\nabla_\xb\cdot \nabla_\zb f_{r_c}(\xb-\zb) a(\xb)\Big]_{\zb=\xb}
\end{split}
\end{equation}
In this integration by part, we have neglected boundary terms which would vanish once applied to a density matrix $\rho_t$ sufficiently well behaved at infinity (which is reasonable for a compact object). As before, the gradient of $f_{r_c}$ evaluated in $0$ vanishes and we are left with:
\begin{equation}
    \int \!\! \upd \xb [\spadesuit] = - 2\, \nabla_\xb^2f_{r_c}(\xb)|_{\xb=0}\, a^\dagger(\zb) a(\zb).
\end{equation}
Carrying the final integration over $\zb$ yields as advertised:
\begin{equation}
    \hat{P}_\text{heat} = \frac{\hbar^2}{2} m \left(-2\,\nabla^2_\xb f_{r_c}(\xb)|_{\xb=0} \right)\int \upd \zb \,a^\dagger(\zb) a(\zb).
\end{equation}
For the CSL model, we have:
\begin{align}
    f^{\text{CSL}}_{r_c}(\xb) &=\frac{\gamma}{2 m^2} \, g_{r_c} * g_{r_c}(\xb)\\
    &=\frac{\gamma}{2 m^2 (\sqrt{4\pi r_c^2})^3}\e^{-\xb^2/(4 r_c^2)}
\end{align}
Hence:
\begin{align}
    -2\,\nabla^2_\xb f^\text{CSL}_{r_c}(\xb)|_{\xb=0} =  \frac{\gamma}{m^2 (\sqrt{4\pi r_c^2})^3} \frac{3}{2r_c^2} = \frac{3 \lambda}{2 m^2r_c^2 }.
\end{align}
Finally, for the CSL model, this gives:
\begin{equation}
P_\text{heat}^{\rm CSL}=\frac{3\lambda \hbar^2}{4 r_c^2 m} N,
\end{equation}
which depends on the quantum state only through the total number of particles.

For the DP model, the regularized kernel $f_{r_c}^\text{DP}$ can easily be evaluated in Fourier space:
\begin{align}
    f_{r_c}^\text{DP}(\xb)&=g_{r_c}*f^\text{DP} * g_{r_c} (\xb)\\
    &= 4 \pi \frac{G}{4\hbar} \int \frac{\upd \kb}{(2\pi)^3} \frac{\e^{-\kb^2 r_c^2}}{\kb^2} \e^{i\kb \cdot \xb}
\end{align}
hence:
\begin{align}
    - \nabla^2_\xb f_{r_c}^\text{DP}(\xb)|_{\xb=0} &=4 \pi \frac{G}{4\hbar} \int \frac{\upd \kb}{(2\pi)^3} \e^{-\kb^2 r_c^2}\\
    &= \frac{G}{8 \sqrt{\pi} r_c^3},
\end{align}
and
\begin{equation}
    P_\text{heat}^\text{DP} = \frac{G\hbar m}{8 \sqrt{\pi} r_c^3} N.
\end{equation}

\end{document}